\documentclass{osa-article}
\articletype{Research Article}
\usepackage{cite}
\usepackage{amsmath, amsfonts, amssymb, wasysym}
\usepackage{siunitx}
\usepackage[caption=false]{subfig}
\usepackage{hyperref}
\usepackage{lipsum, color}
\definecolor{blue}{rgb}{0,0,0.7}
\definecolor{magenta}{cmyk}{0,1,0,0}
\definecolor{darkgreen}{rgb}{0,0.6,0}
\definecolor{purple}{cmyk}{0.5,1,0,0}
\newcommand{\mycomment}[3]{\textcolor{#2}{\textit{(\textbf{#1:} #3)}}}
\newcommand{\marco}[1]{\mycomment{Marco}{blue}{#1}}

\begin{document}

\pagestyle{plain}

\title{Wave-like Properties of Phasor Fields: Experimental Demonstrations}

\author{Syed Azer Reza,\authormark{1} Marco La Manna,\authormark{1} Sebastian Bauer,\authormark{1} and Andreas Velten\authormark{1,2,*}}

\address{\authormark{1}Department of Biostatistics and Medical Informatics, University of Wisconsin -- Madison, Madison, WI 53706, USA\\
\authormark{2}Department of Electrical and Computer Engineering, University of Wisconsin -- Madison, Madison, WI 53706, USA}
\email{\authormark{*}velten@wisc.edu} % Corresponding author
\homepage{\url{https://biostat.wisc.edu/~compoptics/}} %% author's URL, if desired

\begin{abstract*} % Should be +/- 100 words
Recently, an optical meta concept called the Phasor Field (P-Field) was proposed that yields great quality in the reconstruction of hidden objects imaged by non-line-of-sight (NLOS) imaging. It is based on virtual sinusoidal modulation of the light with frequencies in the MHz range. Phasor Field propagation was shown to be described by the Rayleigh-Sommerfeld diffraction integral. We extend this concept and stress the analogy between electric field and Phasor Field. We introduce Phasor Field optical elements and present experiments demonstrating the validity of the approach. Straightforward use of the Phasor Field concept in real-world applications is also discussed. 
\end{abstract*}

%%%%%%%%%%%%%%%%%%%%%%%%%%  body  %%%%%%%%%%%%%%%%%%%%%%%%%%
% --------------------------
\section{Introduction}\label{introduction}
\label{sec:introduction}
% --------------------------
While optical imaging yields great spatial resolution because of the short wavelength, it fails when there is no direct line of sight from the optical detector to the objects to be imaged. However, by utilizing ultrafast pulsed lasers and ultrafast optics, the time-of-flight of the emitted photon pulses can be used to look around corners. This concept, first experimentally demonstrated by Velten et al.~\cite{Velten_12} and often referred to as non-line-of-sight (NLOS) imaging, has since then attracted lots of interest from researchers especially in the fields of optics and computational imaging. Diffuse scattering from visible objects (most times, there is a flat so-called relay wall placed in the scene) enables photons to penetrate into spatial regions not directly visible from the detector. After possibly multiple reflections in the scene, some photons make their way back to the detector. The photons recorded at multiple reflection positions have different travel times depending on the scene geometry. 

By adequately making use of the travel times of all photons, the hidden space can be reconstructed. This geometric reconstruction of the hidden scene is in many cases done by (filtered) backprojection methods \cite{Velten_12,gupta_12} or by methods based on possibly simplified forward light transport models \cite{OToole_18,heide_17,iseringhausen_18}. Most recently, Liu et al.~\cite{liu_18_virtual} demonstrated a wave-like reconstruction approach. The underlying concept, the so-called Phasor Field (P-Field), is based on the fact that for a laser pulse duration in the order of femtoseconds, the acquired time response at each point the detector looks at can be regarded as the impulse response of the scene. Therefore, by convolving the impulse response with a sine wave, we get the response of the scene to such a \textit{virtual} input signal. This essentially makes it possible to consider the relay wall as the aperture of an imaging system, which in turn establishes the analogy between conventional line-of-sight (LOS) and NLOS imaging. The propagation of the modulated virtual light wave is shown to be described with great accuracy by the Rayleigh-Sommerfeld diffraction (RSD) integral. 

In this contribution, we will explore the properties of a \textit{physically} modulated light wave and show that it can be used to look around corners. The vast majority of NLOS studies are based on pulsed light, and while sinusoidally modulated light waves are commonly used for 3D ranging in the direct line-of-sight (LOS) scenario and in distributed media \cite{gupta_15,lin_15,gao_12,lum_18}, there are only very few applications of such waves to NLOS imaging. These few NLOS approaches have in common that they require heavy computation for performing the spatial reconstruction from the acquired measurements \cite{Heide_14, kadambi_16}. The proposed method, by contrast, is based on the physical characteristics of the modulated light, and the reconstruction is therefore just the measurement of an optical phenomenon with almost no calculation involved. These measurements can be performed with a photodiode; the provided electrical signal is both in terms of frequency and signal level in the range of cellphone signals. For this reason, commercial cellphone electronics can be used for its amplification and further processing such as A/D (analog-to-digital) conversion. We will present three experiments conducted to proof the physical analogy between the electric field and P-Field.

We raise the question - is it at all possible to achieve an E-field focus independent of the P-field focus? Or, perhaps, an E-field focus with no P-field focus (this would mean that detected optical irradiance would have no modulation envelope as if the optical source was never modulated to begin with!)? We will present the answer in form of these experiments:
\begin{enumerate}
    \item Repetition of the well-known double-slit experiment, but this time having spatial dimensions on the order of the P-field wavelength. While the unmodulated (direct current, DC) light is scattered everywhere on the screen, the RF (radio frequency) component will show distinct minima and maxima. We therefore have an RF fringe pattern in the absence of an optical fringe pattern (Sec.~\ref{sub_sec:Experiment1}).
    \item A curved diffuser experiment which is mainly comprised of a macroscopic spherical mirror with a diffusing surface. While the light is reflected diffusely (no DC focus), there is an RF focus having the shape of an Airy disk. This means that there is no optical, but a sharp RF focus (Sec.~\ref{sub_sec:Experiment2}). 
    \item A 4-f imaging system realized with four large Fresnel lenses. In this case, all the DC light is focused at one spot, but the magnitude of the RF component at that spot depends on the lens area that is actually illuminated because of interference caused by different path lengths. The illuminated area is controlled using a paraxially placed iris. In total, this experiment realizes sharp optical, but variable RF focus (Sec.~\ref{sub_sec:Experiment3}).
\end{enumerate}
The experiments will be described in detail in Section~\ref{sec_experiments}. The curved diffuser actually functions as a P-field lens, a flat relay wall acts as a P-field mirror, and many more  P-field optical elements can be constructed. The full application potential of this discovery is yet to be explored. In Sec.~\ref{sec:theory}, we briefly recall the phasor field theory. In Sec.~\ref{sec:conclusions}, we provide the result discussion and conclusions.

% --------------------------
\section{P-Field theoretical background}
\label{sec:theory}
% --------------------------
% --------------------------
\subsection{Introduction to Phasor fields}
\label{subsec:theor_intro}
\begin{figure}
 \centering
 \includegraphics[width=2.95in]{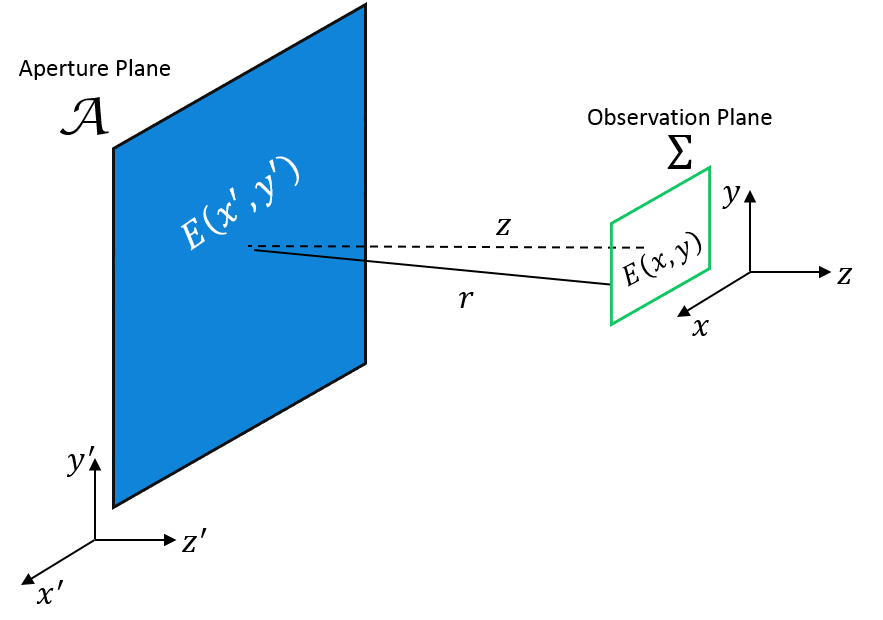}
 \caption{An aperture plane, $\mathcal{A}$, and an observation plane, $\Sigma$, separated by a distance $|r|$.}
 \label{fig:huygens}
\end{figure}

% --------------------------
Let us first provide a brief introduction to the notion of phasor fields as was discussed previously in ~\cite{Reza18j} and ~\cite{Reza18c}. It is well-known that the Huygens' integral is a solution to the scalar wave equation and it describes the Electric field (E-field) at a location $(x,y)$ in a plane $\Sigma$ as a sum of E-field spherical wavelet contributions from all locations $(x',y')$ of another plane $\mathcal{A}$. In the context of imaging, the Huygens' integral
\begin{equation}
    \label{eq:huygens_1}
    E(x,y) = \frac{-1}{j\lambda_\mathrm{E}}\int_{\mathcal{A}}E(x',y')\frac{e^{jK|r|}}{|r|} dx'dy'
\end{equation}
fully explains imaging through an imaging system with $\mathcal{A}$ being the aperture plane and $\Sigma$ defining the image plane as is shown in Fig.~\ref{fig:huygens}. In \eqref{eq:huygens_1}, $|r|=\sqrt{\left(z^2 + x-x' \right)^2 + \left( y-y' \right)^2}$ is the absolute distance between any unique pair of locations $(x',y')$ in $\mathcal{A}$ and $(x,y)$ in $\Sigma$, $z$ is the separation distance between $\mathcal{A}$ and $\Sigma$, $\lambda_\mathrm{E}$ is the E-field (optical) wavelength and $K$ is the E-field wave number expressed as $K = 2\pi/\lambda_\mathrm{E}$. For an amplitude-modulated optical (E-field) signal, we demonstrated that for a 'rough' aperture plane $\mathcal{A}$, the following holds true;
\begin{equation}
    \label{eq:huygens_2}
    \mathcal{P}_\mathrm{Sum} (x,y) \propto \int_{\mathcal{A}}|\mathcal{P}(x',y')|\frac{e^{j\beta|r|}}{|r|} dx'dy'.
\end{equation}
In \eqref{eq:huygens_2}, we denote
\begin{equation}
    \label{eq:P_def}
    \mathcal{P}(x',y') = |\mathcal{P}(x',y')| \frac{e^{j\beta|r|}}{|r|}
\end{equation}
-- the wave-like contributions of the E-field amplitude envelopes from each location $(x',y')$ in $\mathcal{A}$ to a location $(x,y)$ in $\Sigma$ -- as Phasor fields (P-fields) and denote the sum of all P-field contributions at this location $(x,y)$ as $\mathcal{P}_\mathrm{Sum} (x,y)$. In \eqref{eq:P_def}, $\beta$ is the P-field wavenumber expressed in terms of the P-field wavelength $\lambda_\mathrm{P}$ as $\beta = 2\pi/\lambda_\mathrm{P}$ and $|\mathcal{P}(x',y')|$ denotes the magnitude of the P-field contribution from $(x',y')$ in $\mathcal{A}$. As is seen from \eqref{eq:huygens_2}, we observe that P-field contributions from a rough aperture $\mathcal{A}$ add analogously to how E-fields from an aperture add in a typical imaging system described by the Huygens' integral in \eqref{eq:huygens_1}. The summation in \eqref{eq:huygens_2} holds true when the aperture roughness $|\gamma|$ is significantly larger than the E-field wavelength $\lambda_\mathrm{E}$ and much smaller than the P-field modulation wavelength $\lambda_\mathrm{P}$ i.e.,
\begin{equation}
    \label{eq:cond1}
    \lambda_\mathrm{E} \ll |\gamma| \ll \lambda_\mathrm{P}.
\end{equation}
Hence, as was shown in \cite{Reza18j,Reza18c}, this property of P-field summation is ideal for describing an NLOS imaging system in the same way an LOS imaging system is described and modeled using the Huygens' integral because NLOS imaging systems rely on imaging hidden scenes using a relay wall which can be considered as a rough virtual P-field lens aperture in a P-field imaging system.  

% --------------------------
\subsection{An ideal P-field detector}
\label{subsec:P_detector}
% --------------------------

In \eqref{eq:huygens_2}, $\mathcal{P}_\mathrm{Sum} (x,y)$ yields the sum of all slow-varying E-field modulation envelope contributions. As we are more interested in drawing parallels between the Huygens' integral of \eqref{eq:huygens_1} and the P-field integral of \eqref{eq:huygens_2}, we are tempted to only record the magnitude of this P-field sum instead of recording a time-varying signal $\mathcal{P}_\mathrm{Sum} (x,y)$ whose slow temporal changes are recorded by a sufficiently fast photo-detector. Note that this is not the case in E-field imaging as E-field time variations are too fast to be detected by any photo-detector and any typical measurement is only able to measure a time-average optical irradiance $I(x,y) = |E(x,y)|^2$ which is simply a scalar quantity at each location $(x,y)$ in $\Sigma$.

To replicate this in the realm of P-field summation, it is ideally desirable to measure $|\mathcal{P}_\mathrm{Sum} (x,y)|$ instead of simply $\mathcal{P}_\mathrm{Sum} (x,y)$. To achieve this, we define a P-field detector as a combination of two devices. First, there is an AC-coupled photo-diode with sufficiently high electrical bandwidth to measure $\mathcal{P}_\mathrm{Sum} (x,y)$ while removing the DC offset. Second, any electrical component which is able to detect the peak value of $\mathcal{V}_\mathrm{Sum} (x,y)$, the output voltage of the photo-diode which is proportional to $\mathcal{P}_\mathrm{Sum} (x,y)$, to yield $|\mathcal{P}_\mathrm{Sum} (x,y)|$. For all our experiments, we connect the output of an AC-coupled detector to an RF spectrum analyzer. The recorded peak of the spectrum analyzer basically represents the value of the recorded P-field magnitude $|\mathcal{P}_\mathrm{Sum} (x,y)|$.

% --------------------------
\section{Experiments}
\label{sec_experiments}
% --------------------------

For all experiments, the P-field detector was realized using the Menlo systems APD210 photo-detector connected to an Agilent CXA N9000A RF spectrum analyzer which measures $|\mathcal{P}_\mathrm{Sum}|$. As a light source, we were using a laser diode with an optical (E-field) wavelength $\lambda_\mathrm{E}$ of \SI{520}{nm}, while the P-field wavelength was set to roughly \SI{30}{cm} by amplitude modulating the laser diode with a \SI{1}{GHz} sinusoidal signal from a function generator. We could not modulate with a frequency higher than \SI{1}{GHz} due to bandwidth limitations of the optical modulator which we used. We also made simultaneous measurements of optical irradiance with a Thorlabs SM05PD1A power meter measuring the average optical (DC) power.

% --------------------------
\subsection{Experiment 1: P-field double slit interference experiment -- P-field fringe pattern in the absence of an optical fringe pattern}
\label{sub_sec:Experiment1}
% --------------------------

\begin{figure}[ht]
 \centering
 \includegraphics[width=\textwidth]{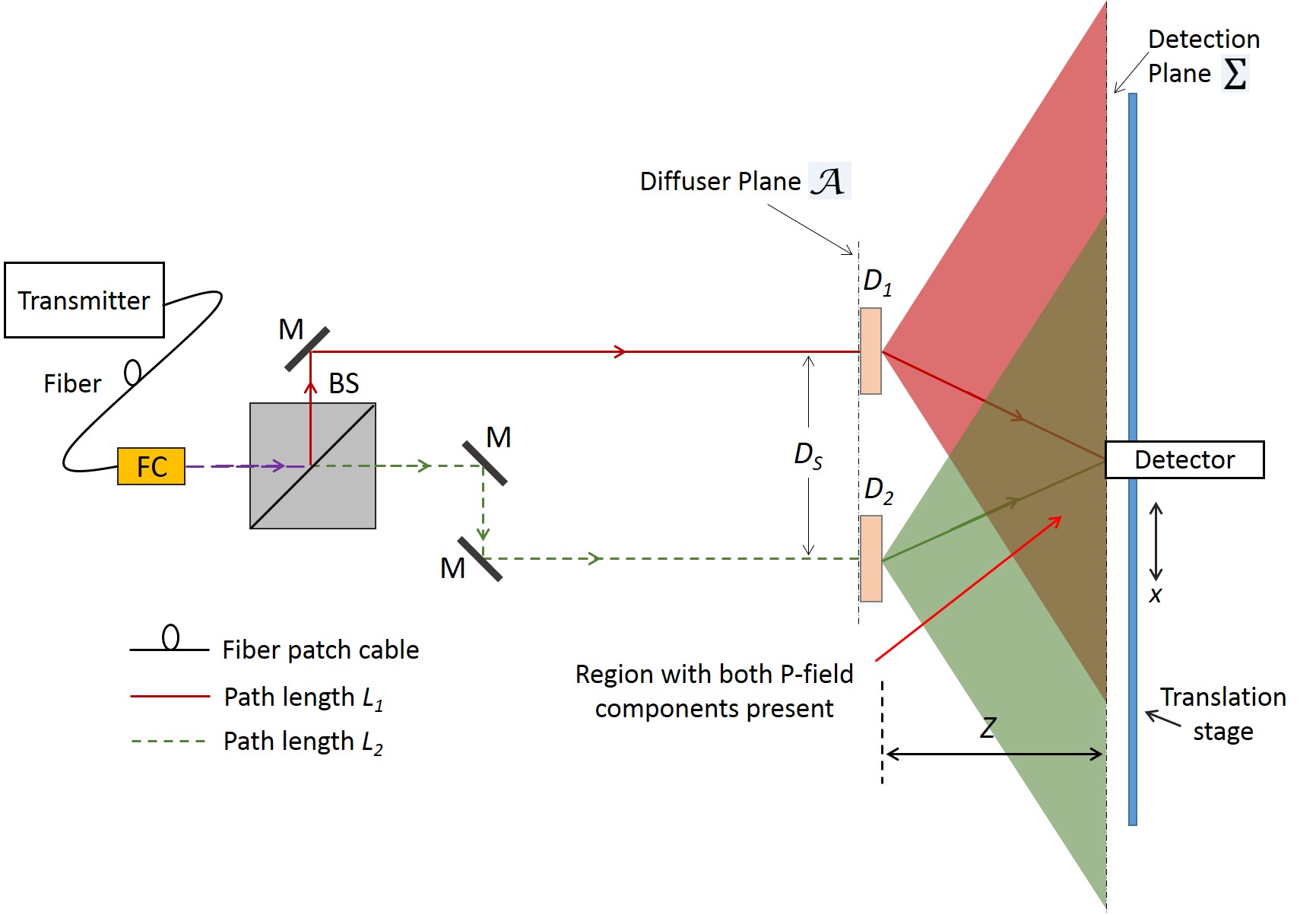}
 \caption{Proposed setup to measure a P-field fringe through a rough aperture.}
 \label{fig:DS_set}
\end{figure}

Our first experiment is analogous to a classical double slit experiment but in the realm of phasor fields. In this experiment, we demonstrate the summation of two P-field contributions and measure a resulting fringe pattern of P-fields  in the detection plane $\Sigma$. The two slits in a classical double slit experiment are replaced by two identical optical diffusers $D_1$ and $D_2$ separated by a distance $D_\mathrm{S}$, as is shown in Fig.~\ref{fig:DS_set}. 

The detection plane $\Sigma$ is located at a distance $z$ from the plane $\mathcal{A}$ containing the two diffusers. Each diffuser randomizes the optical phase of incident photons while having a minimal effect on the phase of the P-field contributions from $D_1$ and $D_2$. These P-field contributions add like waves despite a loss in optical coherence while passing through the rough diffusers. Hence, this setup enacts a P-field double slit experiment where detection with a P-field detector results in the measurement of the sum of two P-field contributions $\mathcal{P}_1$ and $\mathcal{P}_2$ -- one each from $D_1$ and $D_2$ respectively -- analogous to the summation of two E-field contributions in a classical double slit experiment. 

A Gaussian beam from a fiber-coupled laser source exits the fiber through a fiber collimator FC. A 50:50 Beam Splitter (BS) splits the propagating beam into two identical collimated beams, each with exactly half the original beam power. We refer to these two beams as 'Beam 1' and 'Beam 2' and these beams remain collimated until reaching diffusers $D_1$ and $D_2$. Beam 1 and Beam 2 propagate through path lengths L1 and L2 before respective incidence at the two identical diffusers $D_1$ and $D_2$. The experiment was set up such that L1 = L2. Moreover, the P-field amplitude contributions exiting each diffuser are equal i.e., $|\mathcal{P}_2| = |\mathcal{P}_1|$ due to the use of identical diffusers and a 50:50 power beam splitter that delivered an equal optical power split ratio. The expected sum of the two P-field contributions $\mathcal{P}_\mathrm{Sum}$ -- depending on the location $x$ of the detector in $\Sigma$ -- is
\begin{equation}
    \label{eq:output1}
    \mathcal{P}_\mathrm{Sum}(x) = |\mathcal{P}_1(x)|e^{j\phi_1(x)} + |\mathcal{P}_2(x)|e^{j\phi_2(x)} = e^{j\phi_1(x)} \bigg[ |\mathcal{P}_1(x)| + |\mathcal{P}_2(x)| e^{j \Delta\phi_P(x)} \bigg] ,
\end{equation}
where $\phi_2(x)$ has been expressed in terms of the phase difference $\Delta_\mathrm{P} (x) = \phi_2 (x) - \phi_1 (x) $ between the two P-field contributions $\mathcal{P}_1$ and $\mathcal{P}_2$ at detector location $x$. If the P-field detector detects only the magnitude of $\mathcal{P}_\mathrm{Sum}$, the normalized P-field sum $\mathcal{P}_\mathrm{Norm}(x)$ which we theoretically expect to measure  at each location $x$ along the $x$-axis is 
\begin{equation}
    \label{eq:sum_final}
    \left[\mathcal{P}_\mathrm{Norm}(x)\right]_\mathrm{T} = \frac{|\mathcal{P}_\mathrm{Sum}(x)|}{2|\mathcal{P}_1(0)|} = \frac{\bigg| |\mathcal{P}_1(x)| + |\mathcal{P}_2(x)| e^{j \Delta\phi_P} \bigg|}{2|\mathcal{P}_1(0)|},
\end{equation}
where the subscript 'T' denotes the theoretically expected value of $\mathcal{P}_\mathrm{Norm}(x)$. In \eqref{eq:sum_final}, $x = 0$ is the equidistant location from $D_1$ and $D_2$ in the detection plane where $\mathcal{P}_2(0) = \mathcal{P}_1(0)$ because $|\mathcal{P}_2(0)| = |\mathcal{P}_1(0)|$ and $\Delta\phi_\mathrm{P} = 0$. For our measurements, we translated the P-field detector along the $x$-direction and measured the sum of P-field contribution to compute $|\mathcal{P}_\mathrm{Norm}(x)|_\mathrm{M}$ for each detector position where the subscript 'M' denotes the measured value of $\mathcal{P}_\mathrm{Norm}(x)$.

\begin{figure}
 \centering
 \includegraphics[width=\textwidth]{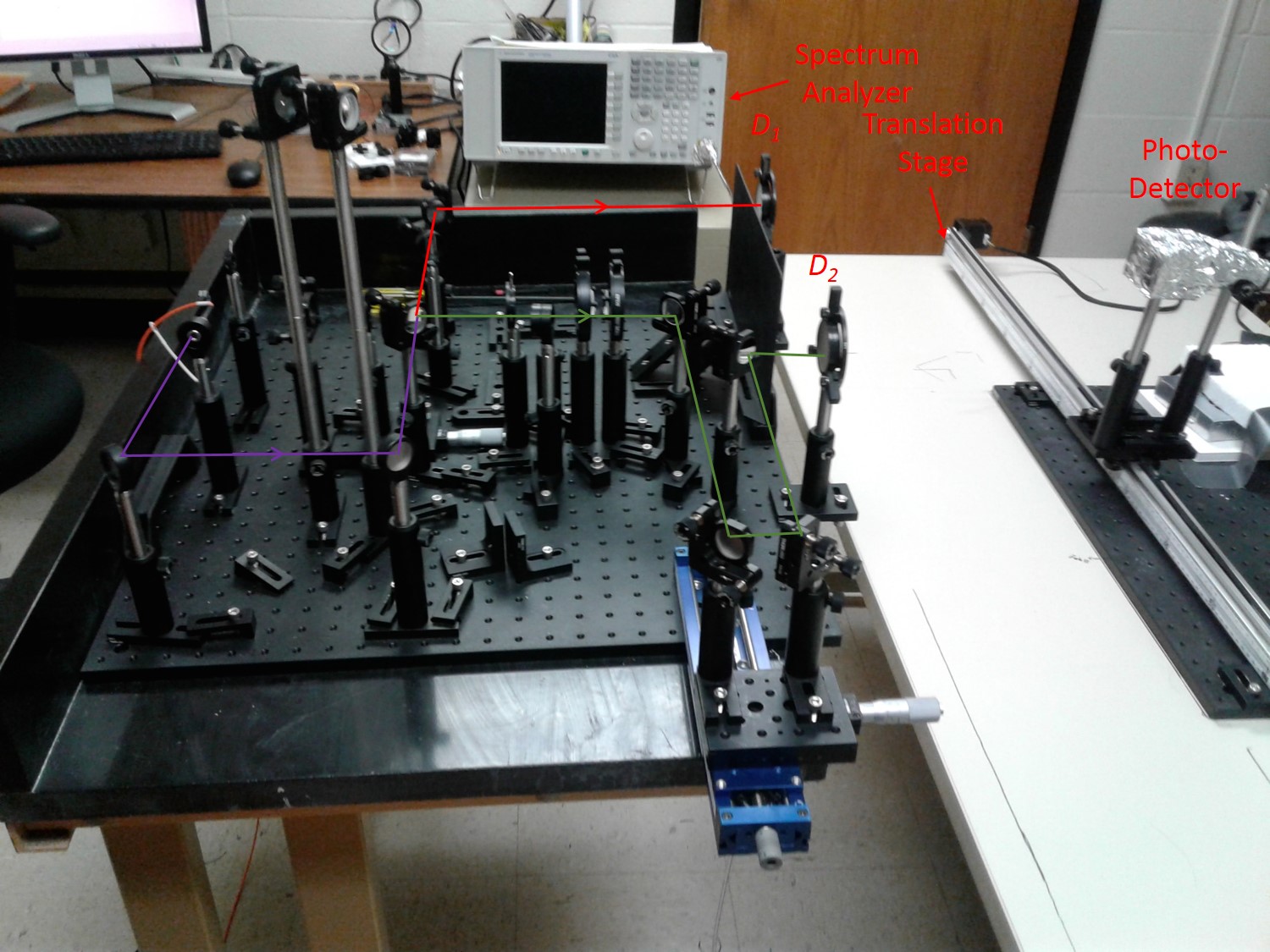}
 \caption{Actual experimental setup to implement system shown in Fig.~\ref{fig:DS_set}}
 \label{fig:DS_set_lab}
\end{figure}

We set the detection plane distance $z = \SI{50}{cm}$ and the diffuser separation distance $D_\mathrm{S} = \SI{36}{cm}$. It also has to be noted that the diffusers used in the experimental setup were part of the the 20DKIT-C3 light-shaping diffusers that do not provide uniform illumination at the $\Sigma$ but instead a Gaussian optical irradiance distribution. We characterized this Gaussian irradiance distribution at $\Sigma$ from $D_1$ and $D_2$ to determine the respective $1/e^2$ Gaussian beam radii $w_{01}$ and $w_{02}$. These measured values were found to be $w_{01} = \SI{35.6}{cm}$ and $w_{02} = \SI{35.7}{cm}$ respectively. A picture of our experimental setup is presented in Fig.~\ref{fig:DS_set_lab}.
%As was done for the first two experiments, a P-field detector was implemented by connecting an AC-coupled Menlo Systems APD210 photo-detector to an Agilent CXA N9000A RF spectrum analyzer to measure $|\mathcal{P}_\mathrm{Sum}(x)|$. 
After the measurement, normalization was performed in post-processing to obtain $\left[\mathcal{P}_\mathrm{Norm}(x)\right]_\mathrm{M}$. %Moreover, as was the case in previous experiments, we used an optical (E-field) wavelength of $\lambda_\mathrm{E} = \SI{520}{nm}$ and the modulation (P-field) wavelength $\lambda_\mathrm{P} = \SI{30}{cm}$. 
\begin{figure}[t]
 \centering
 % 1
 \subfloat[]{
 \label{fig:DS_Meas}
 \includegraphics[width=0.8\linewidth]{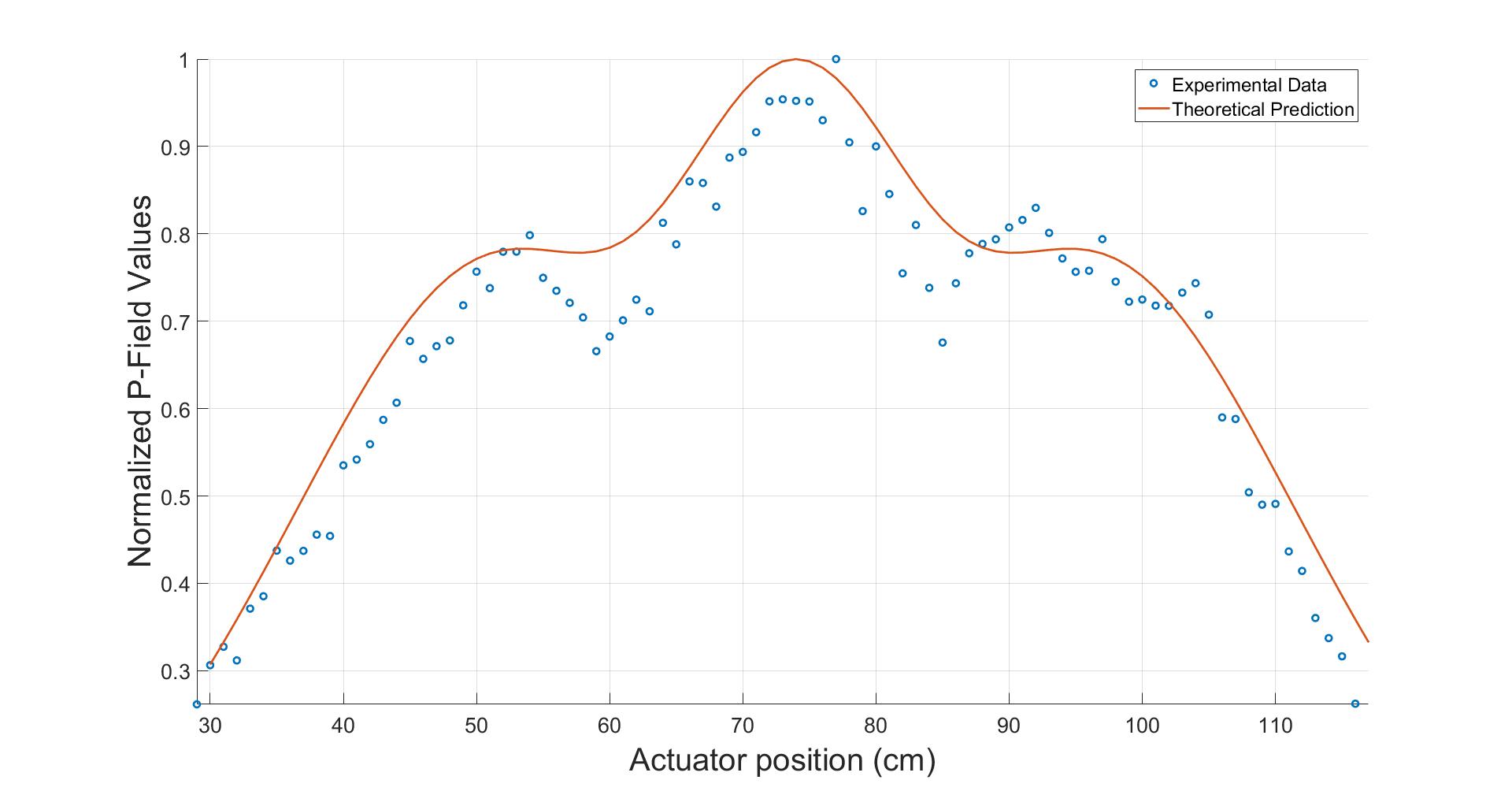}}
 \hfill
  % 2
 \subfloat[]{
 \label{fig:DS_Corr}
 \includegraphics[width=0.8\linewidth]{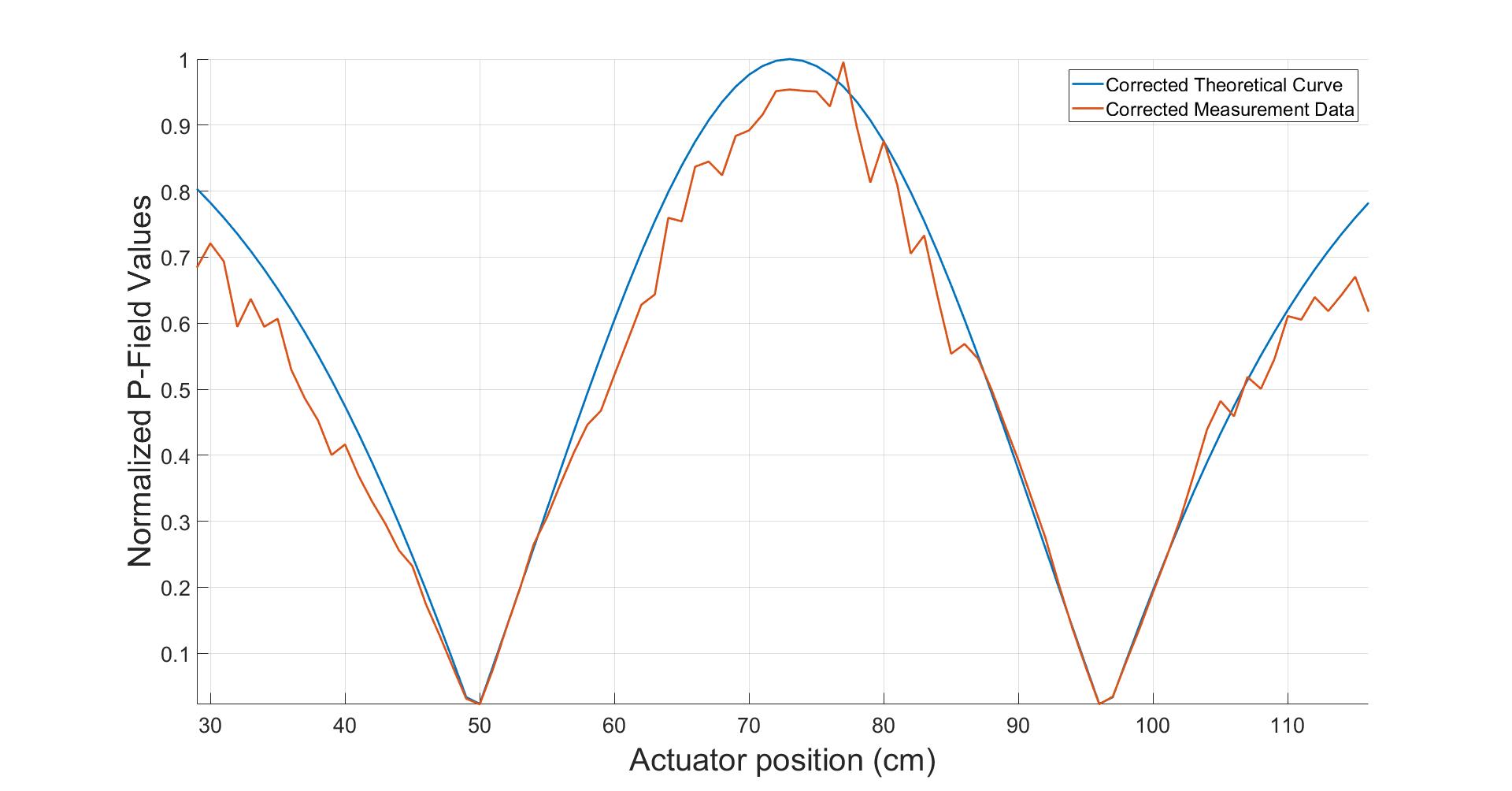}}
% \hfill
 \caption{Plots of (a) theoretically expected $[\mathcal{P}_\mathrm{Norm}(x)]_\mathrm{T}$ values and corresponding experimentally measured P-field normalized sum $[\mathcal{P}_\mathrm{Norm}(x)]_\mathrm{M}$ values for different P-field detector positions $x$ (b) Theoretical and experimental P-field sums $\left[\mathcal{P}_\mathrm{Norm}(x)\right]_\mathrm{M-Uniform}$ corrected for hypothetical uniform optical irradiance contributions from $D_1$ and $D_2$ at $\Sigma$.}
\end{figure}

In Fig.~\ref{fig:DS_Meas}, we plot the theoretically expected as well as experimentally measured normalized P-field sums $\left[\mathcal{P}_\mathrm{Norm}\right]_\mathrm{T}$ and $\left[\mathcal{P}_\mathrm{Norm}\right]_\mathrm{M}$ respectively taking into account the Gaussian irradiance contributions from $D_1$ and $D_2$. There is excellent agreement between measurement data and theoretical predictions. For Gaussian irradiance contributions, instead of uniform contributions, it is evident from Fig.~\ref{fig:DS_Meas} that a complete cancellation of P-field contributions from $D_1$ and $D_2$ is not possible. We instead calculate a theoretical correction factor which provides an equivalent theoretical P-field distribution in the detector plane had there been a uniform irradiance distribution from $D_1$ and $D_2$ instead of a Gaussian one under the same experimental conditions. This correction factor $C(x)$ is computed at every scan location in $\Sigma$ and it is given by
\begin{equation}
    \label{eq:corr1}
    C(x) = \frac{\left[\mathcal{P}_\mathrm{Norm}(x)\right]_\mathrm{T-Uniform}}{\left[\mathcal{P}_\mathrm{Norm}(x)\right]_\mathrm{T-Gaussian}}.
\end{equation}
As is expected, applying the correction factor to $\left[ \mathcal{P}_\mathrm{Norm}(x) \right]_T$, we obtain $\left[\mathcal{P}_\mathrm{Norm}(x)\right]_\mathrm{T-Uniform}$. We also apply the correction factor to the measured data to obtain
\begin{equation}
    \label{eq:corr2}
    \left[\mathcal{P}_\mathrm{Norm}(x)\right]_\mathrm{M-Uniform} = \bigg(\frac{\left[\mathcal{P}_\mathrm{Norm}(x)\right]_\mathrm{T-Uniform}}{\left[\mathcal{P}_\mathrm{Norm}(x)\right]_\mathrm{T-Gaussian}}\bigg)\left[ \mathcal{P}_\mathrm{Norm}(x) \right]_\mathrm{M},   
\end{equation}
which is the equivalent P-field measurement data set had the irradiance from $D_1$ and $D_2$ been uniform under the same experimental conditions. $\left[ \mathcal{P}_\mathrm{Norm}(x) \right]_\mathrm{T}$ and $\left[\mathcal{P}_\mathrm{Norm}(x) \right]_\mathrm{M-Uniform}$ are also plotted in Fig.~\ref{fig:DS_Corr} where, comparing these two curves, it is clearly evident that the peaks, nulls and partial interference in the P-field interference pattern at $\Sigma$ for the corrected measurement dataset closely follow a theoretical P-field fringe pattern for uniform illumination. 

% --------------------------
\subsection{Experiment 2: P-field imaging with a P-field lens -- No optical focus, but P-field focus}
\label{sub_sec:Experiment2}
% --------------------------

\begin{figure}[!ht]
 \centering
 % 1
 \subfloat[]{
 \label{fig:Diff_Theor}
 \includegraphics[width=0.45\linewidth]{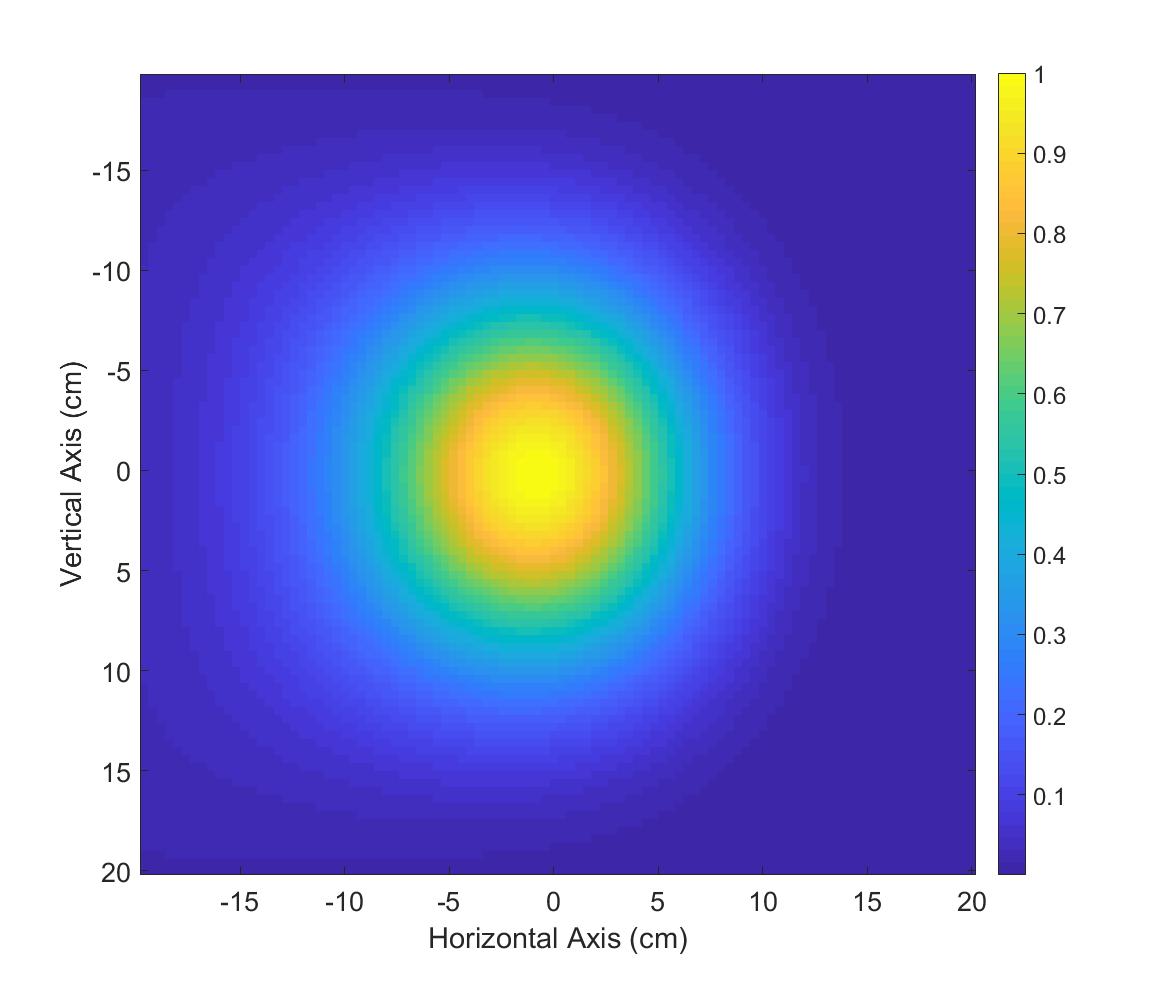}}
 %\hfill
  % 2
 \subfloat[]{
 \label{fig:Diff_Exp}
 \includegraphics[width=0.45\linewidth]{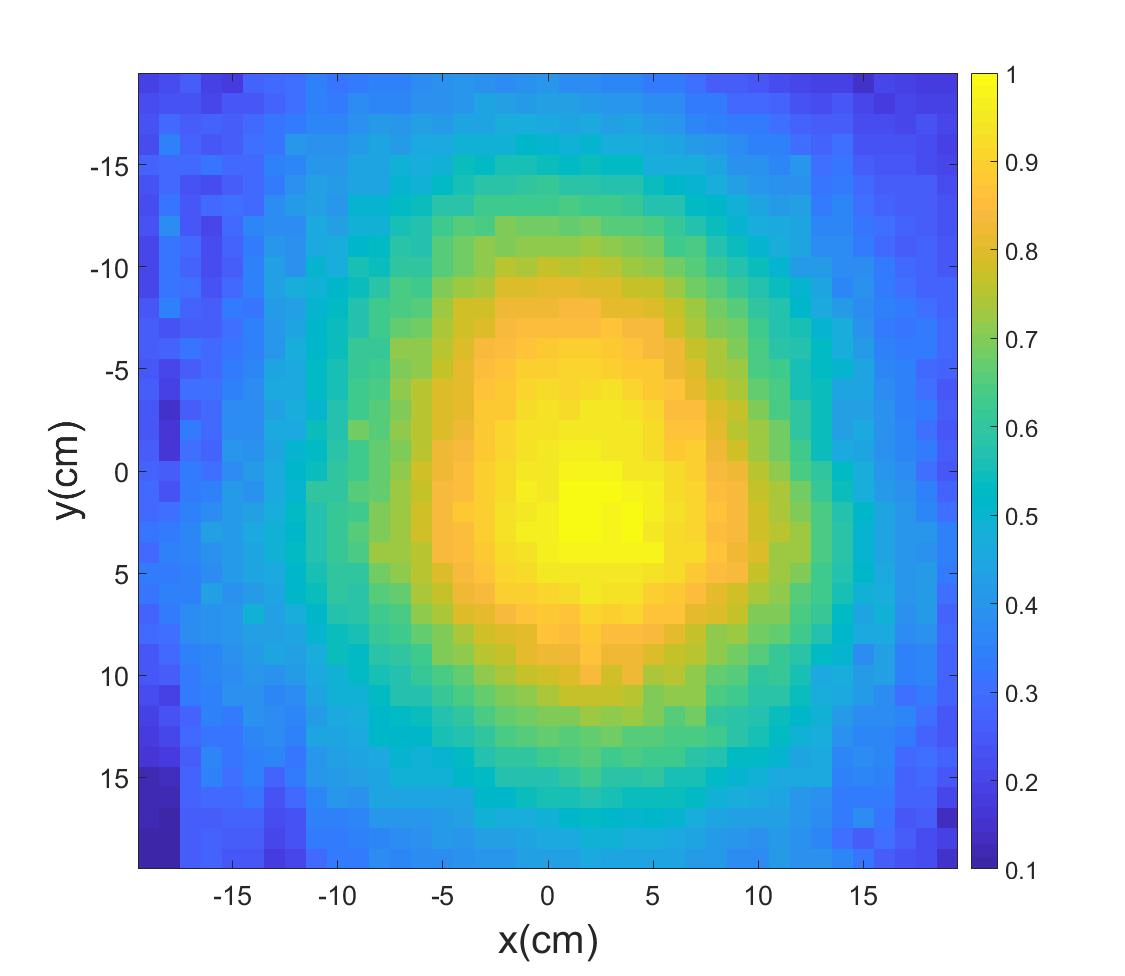}}
 \hfill
   % 3
 \subfloat[]{
 \label{fig:Diff_DC}
 \includegraphics[width=0.70\linewidth]{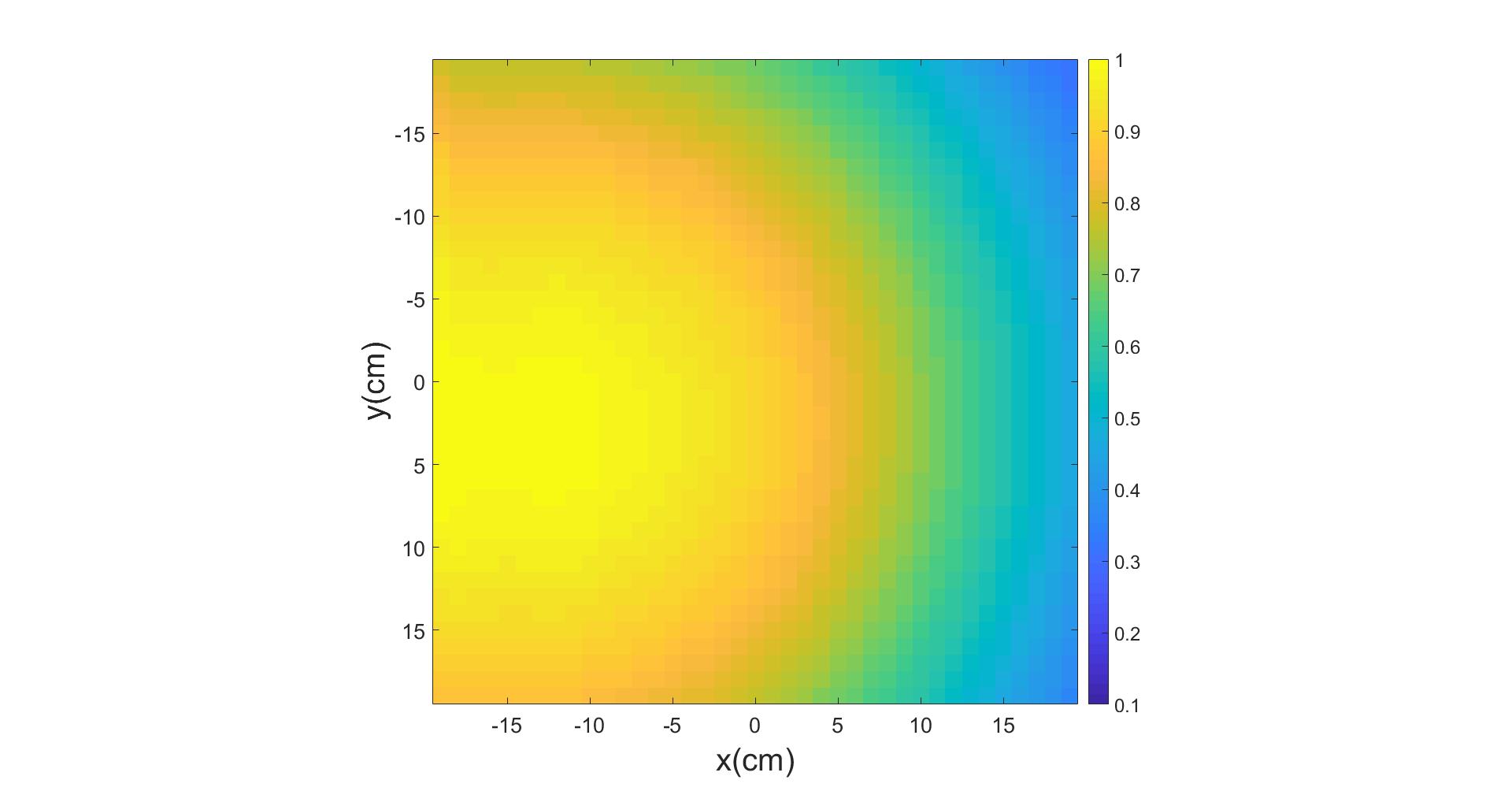}}
 %\hfill
 \caption{Images of (a) theoretically expected P-field focal spot at the image plane, (b) experimentally measured P-field focal spot at the image plane, and (c) experimentally measured optical irradiance in the image plane using a slow photo-detector}
\end{figure}

In this experiment, we demonstrate that P-fields can be focused with the aid of a P-field lens. The operation of a P-field lens is very similar to a conventional lens that enables constructive interference of E-fields from a point object at the location of its focal point. The P-field lens achieves the same for P-fields. The P-field lens can be constructed of a diffuse surface with the correct amount of curvature to obtain a desired focal length value. Moreover, as $\lambda_\mathrm{P} \gg \lambda_\mathrm{E}$, the P-field lens diameter $W$ has to be larger than a conventional E-field lens. The diffuse surface of the P-field lens scatters incident optical photons which illuminate the lens active area. For an amplitude-modulated point light source, we demonstrate that the P-field lens focuses P-fields only and forms a P-field focal spot in the image plane despite the absence of an E-field (or optical) focus and this P-field imaging, for a P-field lens of $f_\mathrm{D}$ follows the well-known lens imaging equation
\begin{equation}
    \label{eq:imag1}
    \frac{1}{f_\mathrm{D}} = \frac{1}{D_\mathrm{Img}} + \frac{1}{D_\mathrm{Obj}}.
\end{equation}
\begin{figure}
 \centering
 \includegraphics[width=\textwidth]{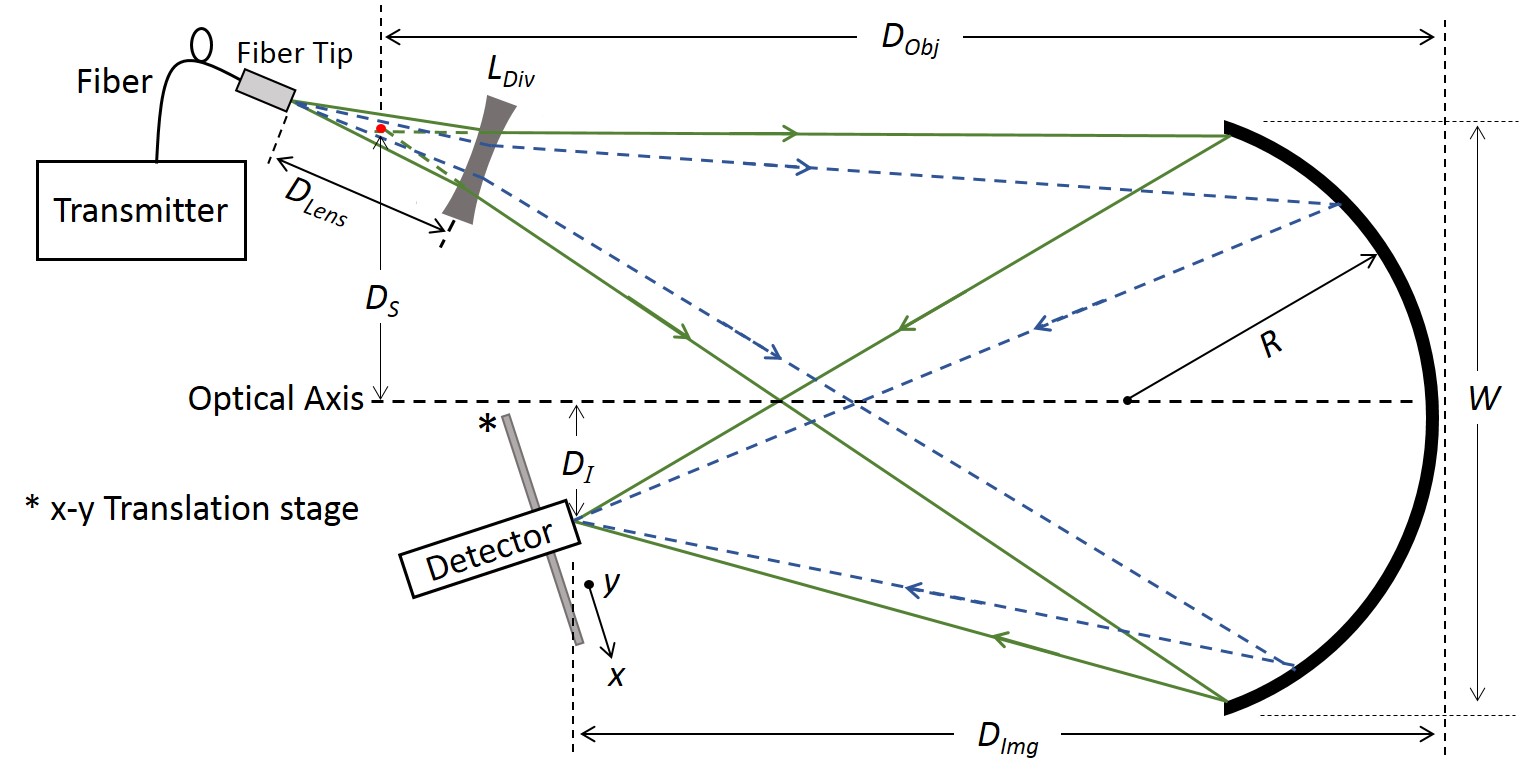}
 \caption{Proposed setup to obtain and measure a P-field focus spot despite no optical (E-field) focus.}
 \label{fig:Diff_set}
\end{figure}
\begin{figure}
 \centering
 \includegraphics[width=\textwidth]{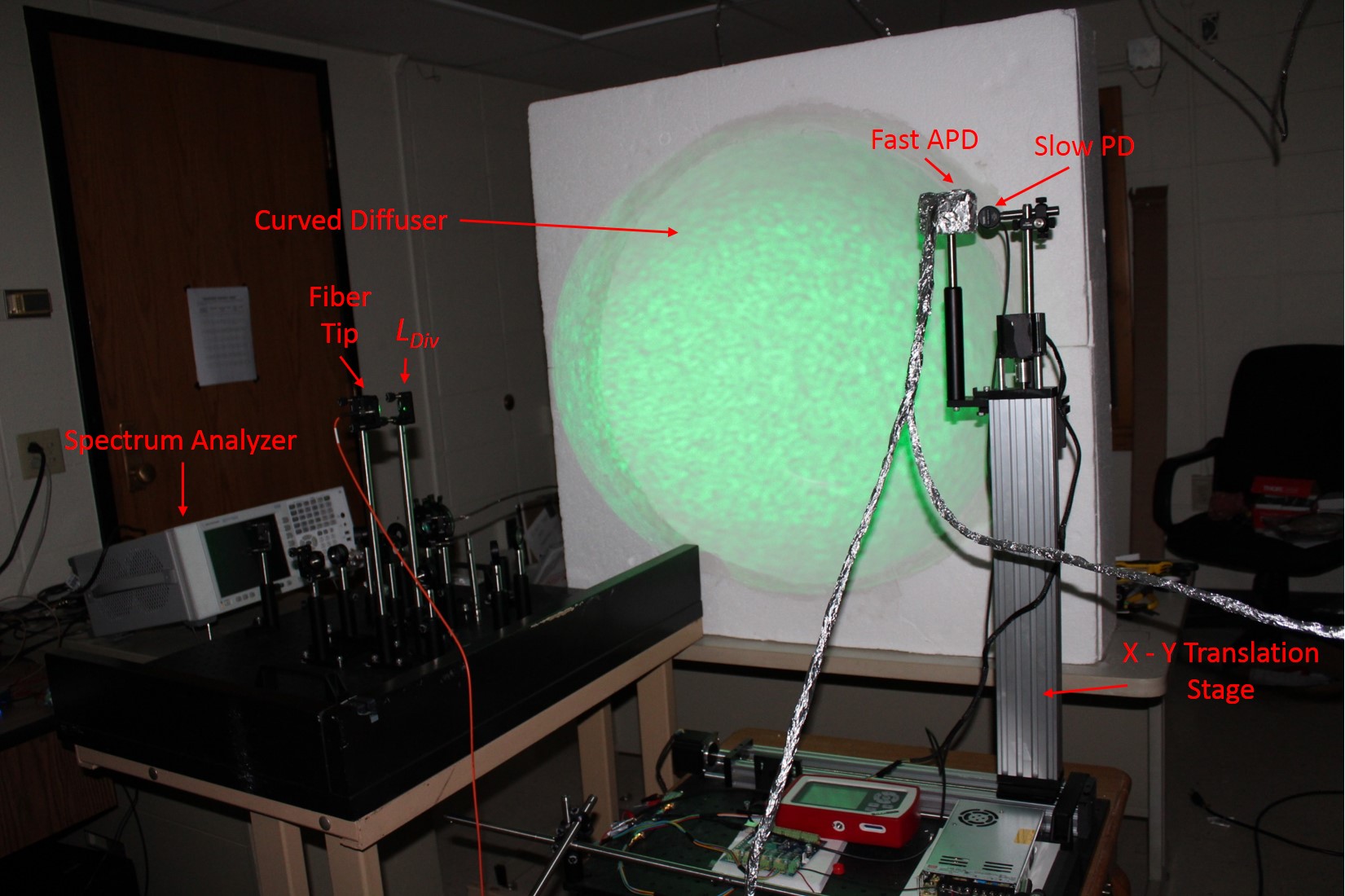}
 \caption{Actual experimental setup to implement system shown in Fig.~\ref{fig:Diff_set}}
 \label{fig:Diff_set_lab}
\end{figure}
This is shown in Fig.~\ref{fig:Diff_set} which illustrates P-field imaging using a curved diffuser. An amplitude-modulated point light source (which can be the tip of an optical fiber) produces a diverging beam. Before incidence on the diffuser, the envelope of each ray defines a phasor-field contribution. Upon incidence on the curved diffuser, the spatial coherence of the optical carrier is severely reduced due to the surface roughness of the diffuser. The roughness $|\gamma|$ of the diffuse surface follows the condition in \eqref{eq:cond1}. Consequently, the phases of different P-field components are minimally affected by the reflection from the curved diffuser and the total P-field phase accumulated can be largely attributed to P-field propagation only. The magnitude of the sum $|\mathcal{P}_\mathrm{Sum}(x,y)|$ of $M \times N$ P-field contributions $|\mathcal{P}_\mathrm{Sum}(x,y)|$ at each image plane location $(x,y)$ can be stated as
\begin{equation}
    \label{eq:diff_sum}
    |\mathcal{P}_\mathrm{Sum}(x,y)| = \bigg|\sum_{i,j=0}^{M,N} \mathcal{P}_{i,j} e^{j\phi_{i,j}}\bigg|.
\end{equation}
In \eqref{eq:diff_sum}, the index $i,j$ denotes the contribution from the $i^{th}$ and $j^{th}$ $x'$ and $y'$ locations at the curved surface of the diffuser. It is only at the focal point location that all phase contributions are equal i.e., $\phi_i = \phi$. Therefore, a maximum value of $|\mathcal{P}_\mathrm{Sum}|$ is recorded at the focal point whereas a radially-symmetric region of partial  P-field interference surrounds the focal point resulting in a P-field spot similar to a diffraction-limited spot in the case of E-field imaging with a lens. The entire P-field spot (including radially symmetric regions of partial interference around the focal spot) is very similat to an Airy spot encountered and described in conventional imaging \cite{goodman2005introduction}. We performed simulations, where we sum all P-field contributions from a point object and compute the magnitude of this sum at every location in the image plane (using \eqref{eq:diff_sum}). The simulated P-field spot in the image plane is plotted in Fig.~\ref{fig:Diff_Theor}. 

For the actual experiment, we constructed a curved diffuser by stacking various styrofoam layers with holes of different radii, gluing these layers together and smoothing the resulting diffuser surface using gypsum paste. This diffuser was designed and constructed for a desired radius of curvature value of $R = \SI{50}{cm}$ and, consequently, a diffuser focal length $f_\mathrm{D} = \SI{25}{cm}$. Additionally, the diameter of the curved diffuser is $W = \SI{82}{cm}$. The surface quality of the diffuser is better than $\lambda_\mathrm{P}/30$. Amplitude-modulated light from a fiber-coupled laser diode (LD) exits the fiber and diverges rapidly. To ensure that the diffuser is completely illuminated with a minimal number of photons outside the curved region of the diffuser, we place a spherical concave lens $L_\mathrm{Div}$ of focal length $f_\mathrm{Div}$ to enhance beam divergence. The curved diffuser is placed at a distance $D_\mathrm{Obj}$ from the virtual source point $P$ behind $L_\mathrm{Div}$. As per \eqref{eq:imag1}, a P-field image spot is expected to form at a distance of roughly $D_\mathrm{Img}$ from the diffuser. As the expected P-field spot is much larger than the active area of the photo-detector (which is responsible for capturing photons), the detector is scanned horizontally and vertically in small steps to measure $|\mathcal{P}_\mathrm{Sum}(x,y)|$ at each image plane location $(x,y)$. For clarity, a picture of our experimental is shown in Fig.~\ref{fig:Diff_set_lab}. The orthogonally mounted large translation stages, which enable scanning of the image plane, are also marked in Fig.~\ref{fig:Diff_set_lab}. 

As was the case with the first experiment, the E-field wavelength of the optical source was chosen to be $\lambda_\mathrm{E} = \SI{520}{nm}$, the P-field wavelength was set to $\lambda_\mathrm{P} \approx \SI{30}{cm}$, the distances $D_\mathrm{Obj}$ and $D_\mathrm{Img}$ were set to $\SI{69.3}{cm}$ and $\SI{43}{cm}$ respectively. Moreover, referring to Fig.~\ref{fig:Diff_set}, distances $D_S$ and $D_I$ were set to $\SI{40}{cm}$ and $\SI{22.5}{cm}$ respectively. Also, $L_\mathrm{Div}$ was a concave lens with $f_\mathrm{Div} = \SI{-30}{mm}$, placed at a distance of $D_\mathrm{Lens} = \SI{5}{cm}$ from the fiber tip. As was mentioned earlier, our P-field detector design remains consistent for all experiments with an AC-coupled Menlo Systems APD210 photo-detector connected to an Agilent CXA N9000A RF spectrum analyzer in order to measure $|\mathcal{P}_\mathrm{Sum}(x,y)|$. 

Measured $|\mathcal{P}_\mathrm{Sum}(x,y)|$ values for all scan locations of the image plane are plotted in Fig.~\ref{fig:Diff_Exp}. We observe a P-field focal spot with a peak in the center and a radially-symmetric amplitude drop-off. The P-field spot is slightly wider than what theory predicts. This can be due to potential difference between the ideal location where the image plane was perceived to be and where it actually was. Also, for our simulations, we assumed an ideal uniform illumination of the diffuser - which was clearly not the case as our illumination from the point source was Gaussian.

The slow SM05PD1A power meter was positioned in a near-confocal configuration with respect to the fast Menlo Systems APD210 photo-detector. The measured irradiance distribution normalized to the highest measured irradiance value in the image plane is presented in Fig.~\ref{fig:Diff_DC} and shows that the measured P-field focal spot is not an artifact of the E-field distribution over the image plane.    

% --------------------------
\subsection{Experiment 3: Imaging through Fresnel lenses -- Sharp optical focus with a variable P-field focus}
\label{sub_sec:Experiment3}
% --------------------------

The objective of this experiment is to demonstrate the formation of a tightly focused optical (E-field) image spot of a modulated optical 'point-like' source (due to constructive E-field interference) while the modulating envelope is either completely, partially or barely detectable as a result of a constructive, partial or destructive interference of P-fields. Hence, this experiment aims to validate our claim that we could possibly realize a scenario where a diffraction-limited optical (E-field) focus is observed in an image plane $\Sigma$ of an optical system whilst concurrent P-field focus is partially or completely absent at the location of E-field focus.
\begin{figure}
 \centering
 \includegraphics[width=\textwidth]{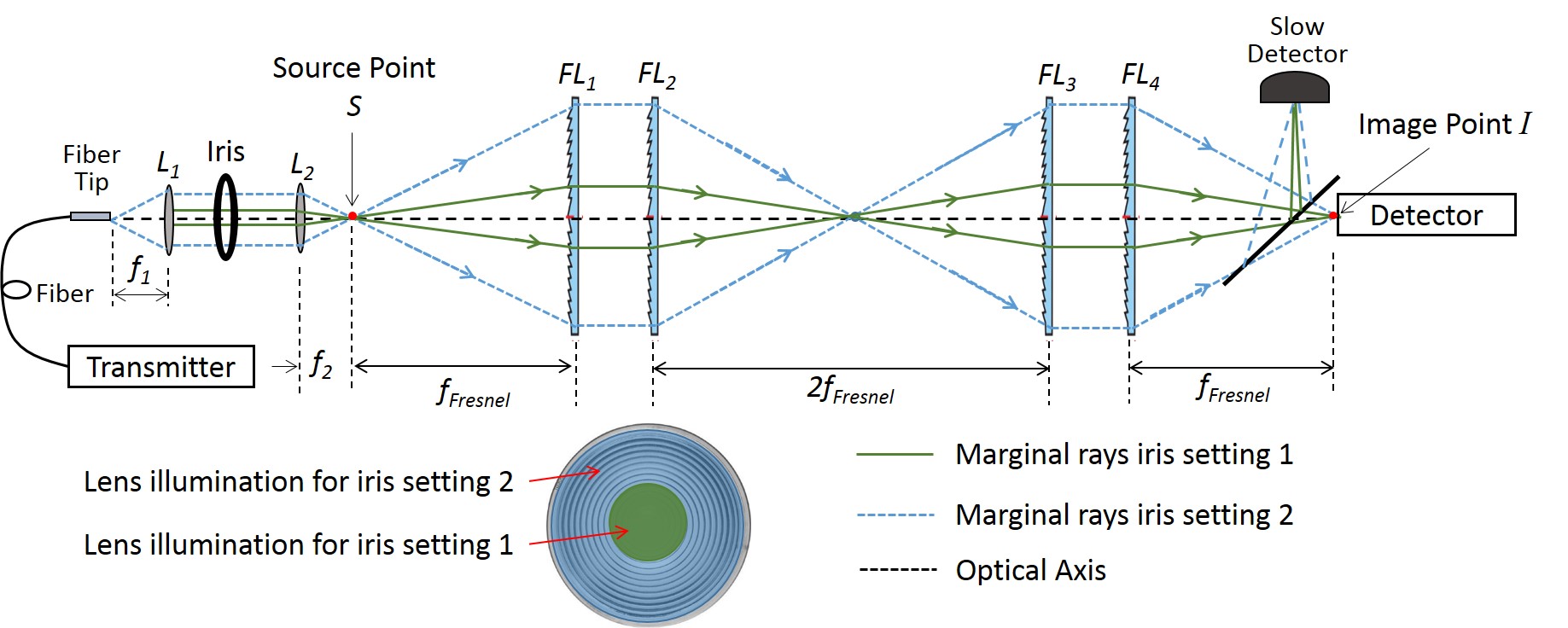}
 \caption{Setup to demonstrate optical focus with varying P-field focus}
 \label{fig:Fres_set}
\end{figure}
For this purpose, we propose the experimental setup of Fig.~\ref{fig:Fres_set} where we construct a 4-f imaging system which creates a diffraction-limited point image $I$ of a point light source $S$ in the image plane $\Sigma$. As is expected, this imaging system enforces constructive interference of various E-field contributions from $S$ at the location $I$ where we observe a sharp optical focus and refer to it as the image point.

If four conventional spherical lenses are used in the setup of Fig.~\ref{fig:Fres_set}, the number of phase cycles added by the four lenses to each of the E-field contributions from the point source counter-balances the differences in the accumulated phases due to propagation outside of the lenses (i.e. in Fig.~\ref{fig:Fres_set} for propagation from source point to lens 1, between lens 1 and lens 2, lens 2 and lens 3, lens 3  to lens 4 and from lens 4 to the image point). Hence imaging is achieved by conventional spherical lenses through optical path balancing of different E-field components from a point source and it allows for perfectly constructive E-field (optical) interference at the image point. Moreover, for an amplitude modulated optical point source, the lens also counter-balances the path length differences of various P-field contributions. Therefore, for any ideal imaging system that uses conventional spherical lenses (including the classical 4-f imaging system), the location of highest P-field constructive interference is the same as the optical image point $I$ which signifies the location of the highest constructive interference of E-fields. 

We now draw the attention to optical lenses which do not perform achieve imaging using the principle of optical path length balancing. Fresnel lenses are one such type of lenses which achieve imaging by imparting the correct amount of phase to each E-field contribution instead of fully balancing optical paths i.e. E-field contributions can constructively interfere if each contribution accumulates a different number of complete $2\pi$ phase cycles but the same phase residual $\phi$ for each of the contribution i.e.
\begin{equation}
    \label{eq:E_sum}
    \theta_i = 2\pi N_i + \phi,
\end{equation}
where $\theta_i$ is the total phase accumulated by the $i^{th}$ E-field contribution and $N_i$ is an integer number of complete $2\pi$ phase cycles accumulated by this $i^{th}$ E-field contribution. Constructive interference is possible in this manner due to the modulo $2\pi$ nature of time harmonic functions which form the bases of a Fourier series which any signal can be expressed as. 

Hence Fresnel lenses are thin as these lenses only aim to achieve phase residual ($\phi$) balancing instead of a full-on optical path balancing. If the average thickness $\langle T \rangle$ of Fresnel lenses is much less than the P-field wavelength, i.e. $\langle T \rangle \ll \lambda_\mathrm{P}$, it is entirely possible that propagation through a series of Fresnel lenses has almost no effect on the phase that each P-field contribution accumulates. Hence all P-field contributions at the E-field image point in $\Sigma$ could have dissimilar phase accumulations which would be primarily due to propagation between $S$ and $I$ with almost no phase contribution from propagation within the Fresnel lenses.    

In other words, this clear distinction between any conventional spherical lens and a Fresnel lens allows for a non-identical phase accumulation for all P-field contributions from the source to the image point despite an identical phase accumulation for all E-field contributions. This results in constructive interference of E-field contributions and a partial interference of the P-field contributions. 

The 4-f imaging system in Fig.~\ref{fig:Fres_set} uses four identical Fresnel lenses, labeled as $FL_1$, $FL_2$, $FL_3$ and $FL_4$, each with a focal length of $f_\mathrm{Fresnel}$ and diameter $D_\mathrm{Fres}$. Amplitude-modulated light from a fiber-coupled laser diode (LD) exits through the fiber tip at an fixed divergence angle $\alpha$. A spherical lens $L_1$ with a focal length $f_1$ is placed at a distance $f_1$ from the fiber tip such that the beam exiting $L_1$ is collimated. The choice of the focal length and size of $L_1$ depends on the numerical aperture of the fiber and the desired collimated beam size after $L_1$. The beam then passes through the second spherical lens $L_2$ of an equal diameter as $L_1$. $L_2$ is placed at a distance $D_\mathrm{Sep}$ from $L_1$ and the choice of its focal length $f_2$ depends on the desired divergence angle which fully illuminates the first Fresnel lens $FL_1$. From Fig.~\ref{fig:Fres_set}, $L_2$ first focuses the incident collimated beam to a point $S$ located at a distance $f_2$ from $L_2$. From there on, this beam diverges to fully illuminate $FL_1$ placed at a distance $f_\mathrm{Fres}$ from $S$ to ensure that the much larger beam which exits $FL_1$ is collimated again. The beam then passes through a series of Fresnel lenses $FL_2$, $FL_3$ and $FL_4$. In-line with how a classical 4-f imaging system is typically implemented, the choice of separation distance between lens pairs $FL_1$/$FL_2$ and $FL_3$/$FL_4$ is arbitrary while the separation distance between $FL_2$ and $FL_3$ is set to $2f_\mathrm{Fres}$ and between $FL_4$ and the receiver to $f_\mathrm{Fres}$. This lens arrangement results in a collimated between $FL_3$ and $FL_4$ and the formation of an image spot at location $I$ which is at a distance $f_\mathrm{Fres}$ from $FL_4$. 

Different modulated optical rays from $S$ propagate through the system and arrive at $I$. The optical carrier (E-field) phase accumulation for all rays is identical resulting in the formation of an optical focal spot at $I$ while the P-field phase accumulation associated with the modulation envelope of each ray is dissimilar for propagation between $S$ and $I$. The path length differences between P-field contributions from the fiber tip to source point $S$ are considered negligible if the diameters of $L_1$ and $L_2$ as well as their respective focal lengths $f_1$ and $f_2$ are chosen to be much shorter than the P-field wavelength $\lambda_\mathrm{P}$. For complete illumination of each Fresnel lens (note that for the 4-f imaging system in Fig.~\ref{fig:Fres_set} the incident irradiance distribution is identical at each Fresnel lens and therefore it is not important to mention which lens irradiance distribution we are speaking of), the magnitude of the sum $|\mathcal{P}_\mathrm{Sum}(D_\mathrm{Fres})|$ of all P-field contributions detected by the P-field detector at $I$, is simply expressed as a function of Fresnel lens diameter $D_\mathrm{Fres}$ as
\begin{equation}
    \label{eq:P_sum2}
    |\mathcal{P}_\mathrm{Sum}(D_\mathrm{Fres})| = \bigg|\sum_{i=0}^{i_\mathrm{Max}(D_\mathrm{Fres})} \mathcal{P}_i\bigg| =  \bigg|\sum_{i=0}^{i_\mathrm{Max}(D_\mathrm{Fres})} |\mathcal{P}_i|e^{j\phi_i}\bigg|.
\end{equation}
In \eqref{eq:P_sum2}, $\phi_i$ is the phase accumulated between $S$ and $I$ by the $i^{th}$ P-field contribution $\mathcal{P}_i$ of amplitude $|\mathcal{P}_i|$. As the 4-f imaging system in Fig.~\ref{fig:Fres_set} is radially symmetric, $\mathcal{P}_i = |\mathcal{P}_i|e^{j\phi_i}$ denotes the $i^{th}$ contribution of the modulation envelope of all photons in the $FL_1$ plane which are located at a radial distance $r_i$ from the center of $FL_1$ (consequently true for all other Fresnel lenses as well). In other words, the P-field contribution with a specific amplitude $|\mathcal{P}_i|$ and phase $\phi_i$ is produced by modulated photons located in a thin ring of radius $r_i$ centered at the center of $FL_1$. The amplitude $|\mathcal{P}_i|$ of the P-field contribution $\mathcal{P_i}$ is determined by the number of photons incident at  $r = r_i$ and its phase $\phi_i$ relative to the shortest on-axis path through the system is given by
\begin{equation}
    \label{eq:p_phase}
    \phi_i = \left( \frac{2\pi}{\lambda_\mathrm{P}} \right) 4\left( \sqrt{r_i^2 + f_\mathrm{Fres^2}} -f_\mathrm{Fres}\right) = \left( \frac{8\pi f_\mathrm{Fres}}{\lambda_\mathrm{P}} \right) \bigg(\sqrt{\frac{r_i^2}{f_\mathrm{Fres}^2}+1} -1 \bigg).
\end{equation}
To demonstrate a comprehensive P-field summation behavior, a mechanically tunable iris is placed between $L_1$ and $L_2$ and it is centered at the collimated beam present in that location. The collimated beam passes through the optical iris and the area of the circular optical illumination of each of the Fresnel lenses is altered for each different setting of the iris opening. The iris is chosen such that when its fully open it allows the collimated beam between $L_1$ and $L_2$ to propagate through unchopped resulting in the subsequent illumination at each Fresnel lens to remain unchopped. 

Moreover, to demonstrate a comprehensive range of P-field contructive and destructive interference at $I$, the diameter $D_\mathrm{fres}$ of each Fresnel lens was chosen such that for the iris fully open (i.e. complete Fresnel lens illumination), the maximum path length difference $\Delta_L$ between P-field contributions with the shortest and longest propagation paths $L_\mathrm{Min}$ and $L_\mathrm{Max}$ between $S$ and $I$ was greater than or equal to the P-field wavelength $\lambda_\mathrm{P}$ i.e.,
\begin{equation}
    \label{eq:phase_accum}
    \Delta_L = L_\mathrm{Max} - L_\mathrm{Min} \geq \lambda_\mathrm{P}.
\end{equation}
This ensures that each P-field contribution $\mathcal{P}_i$ with a respective phase $\phi_i$ at $I$ in the range $0 \leq \phi_i \leq \pi$ has a conjugate P-field phase contribution in the range $\pi \leq \phi_i \leq 2\pi$ to destructively interfere with it. This choice of $D_\mathrm{Fres}$ ensures that we are able to observe the largest possible constructive and destructive interference of P-field contributions as well as all intermediate interference states for different iris settings. Therefore
\begin{equation}
    \label{eq:phase_accum2}
     \Delta_L = 4 f_\mathrm{Fres}\bigg(\sqrt{\frac{D^2}{f_\mathrm{Fres}^2}+1} -1 \bigg) \geq \lambda_\mathrm{P},
\end{equation}
\begin{equation}
    \label{eq:phase_accum3}
    \implies D_\mathrm{Fres} \geq \frac{\lambda_\mathrm{P}}{2} \sqrt{\frac{1}{4} + \frac{2 f_\mathrm{Fres}}{\lambda_\mathrm{P}}}.
\end{equation}
With the iris present within the setup, the sum of P-field contributions $|\mathcal{P}_\mathrm{Sum}|$ detected is a function of the optical illumination radius $R$ corresponding to an iris opening radius of $R$. It is expressed as
\begin{equation}
    \label{eq:P_sum1}
    |\mathcal{P}_\mathrm{Sum}(R)| = \bigg|\sum_{i=0}^{i_\mathrm{Max}(R)} |\mathcal{P}_i|e^{j\phi_i}\bigg|,
\end{equation}
where $i_\mathrm{Max}$ is the maximum allowable value of index $i$ for any particular iris opening. 

\begin{figure}
 \centering
 \includegraphics[width=\textwidth]{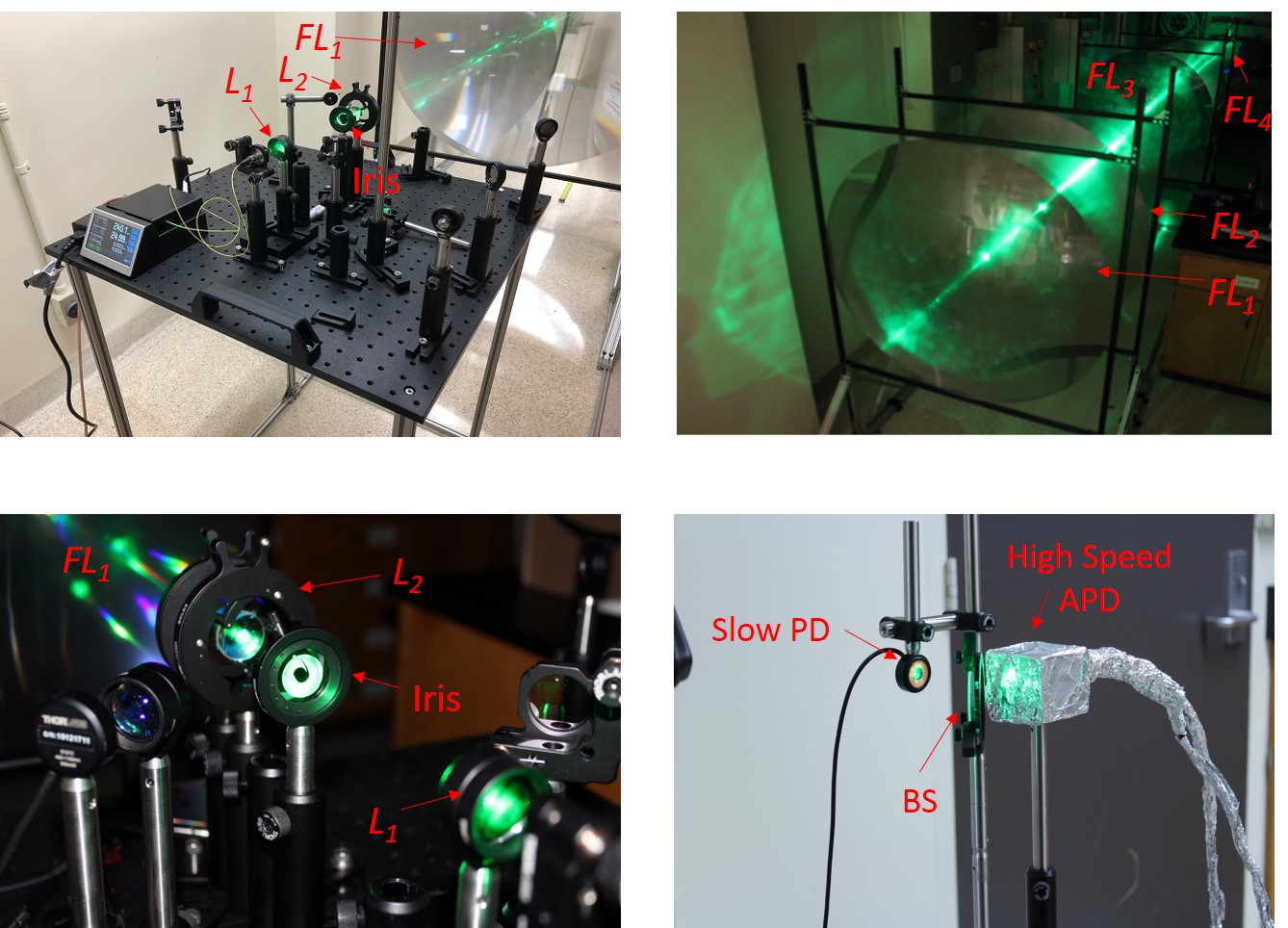}
 \caption{Actual experimental setup to implement system shown in Fig.~\ref{fig:Fres_set}}
 \label{fig:Fres_set_lab}
\end{figure}

For the actual experiment, the photo-detector is located at $I$. P-field measurements $|\mathcal{P}_\mathrm{Sum}(R)|$ were recorded at various iris radius ($R$) settings. These measurements were normalized by the highest recorded value of $|\mathcal{P}_\mathrm{Sum}(R)|$ which we denote as ${|\mathcal{P}_\mathrm{Sum}|}_\mathrm{Max}$.
For the P-field frequency $\Omega = \SI{1}{GHz}$, we chose four Fresnel lenses CP1300-1100 by FresnelFactory.com, each with $D_\mathrm{Fres} = \SI{1100}{mm}$ and a focal length $f_\mathrm{Fres} = \SI{1300}{mm}$. This choice of using large Fresnel lenses was governed by the condition in \eqref{eq:phase_accum3}, which these lenses satisfy when used in a 4-f imaging system shown in Fig.~\ref{fig:Fres_set}. Lens $L_1$ used in the experiment was a 1-inch spherical lens with $f_1 = \SI{50}{mm}$ which resulted in a \SI{1.1}{cm} null-to-null diameter of the collimated beam after $L_1$. Instead of using a single lens $L_2$, we replaced it with a lens pair comprising of two 1-inch spherical lenses separated by a distance of \SI{7}{mm}. The first lens had a focal length of \SI{30}{mm} while the second lens had a focal length of \SI{50}{mm}. This allowed us to overshoot the illumination of $FL_1$ with the iris fully open and illuminate the lens more uniformly and ensure full lens illumination at a particular opening radius of the iris. The resulting beam illumination at $FL_1$ was Gaussian with a $1/e^2$ radius of $w_0 \approx \SI{84}{cm}$ which was measured separately. Pictures of the actual experimental setup are presented in Fig.~\ref{fig:Fres_set_lab}.

\begin{figure}[htbp]
 \centering
 % 1
 \subfloat[]{
 \label{fig:Fres_AC}
 \includegraphics[width=0.7\linewidth]{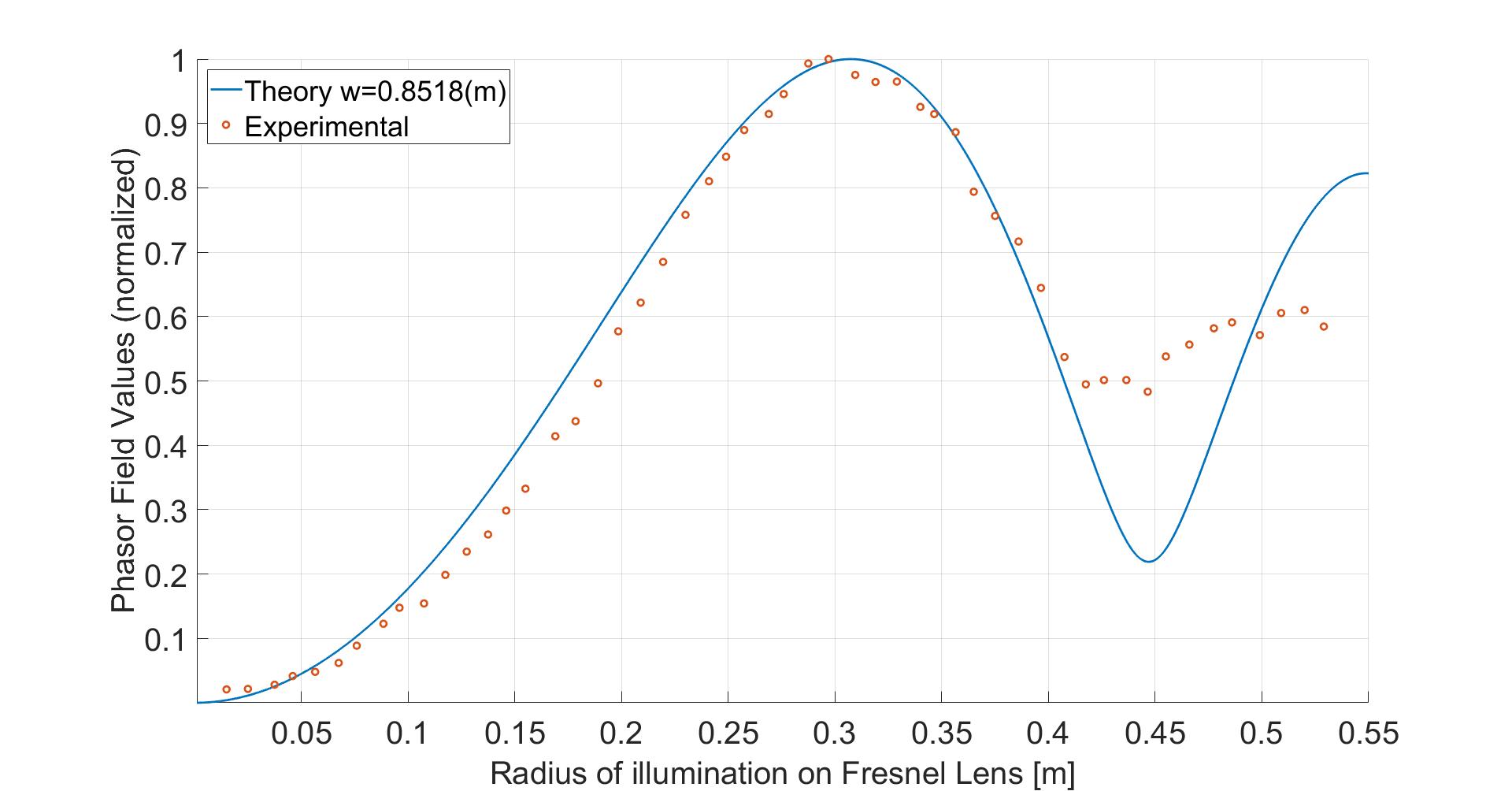}}
 \hfill
  % 2
 \subfloat[]{
 \label{fig:Fres_DC}
 \includegraphics[width=0.7\linewidth]{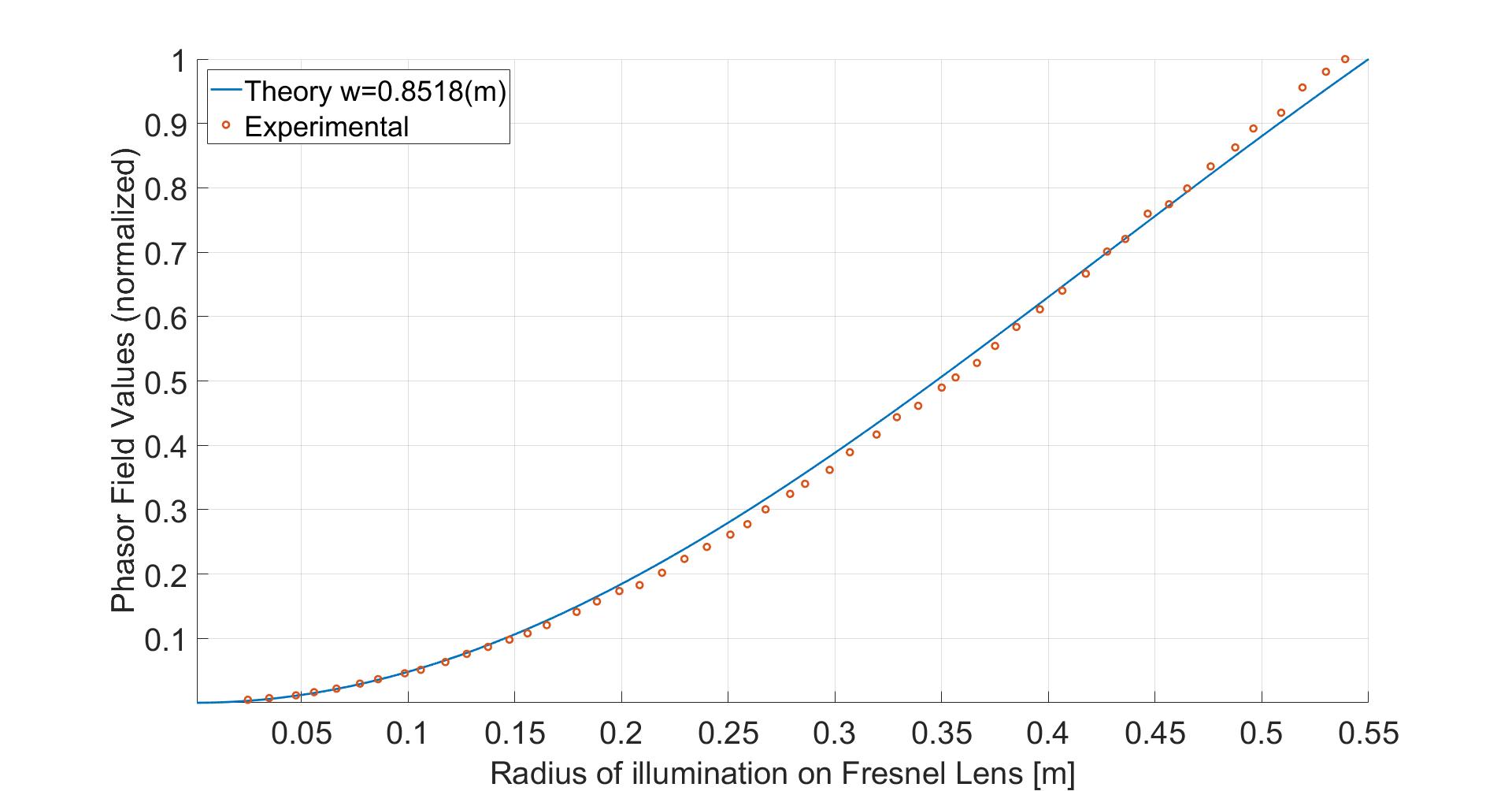}}
 \hfill
 \caption{Plot showing (a) comparison between expected theoretical P-field summation $\mathcal{P}_\mathrm{Norm}(R)$ for different $FL_1$ illumination radius $R$ values and experimentally measured data, and (b) a comparison of $\left(P_\mathrm{DC-Norm}\right)_\mathrm{Theoretical}$ to experimentally measured values of $\left(P_\mathrm{DC-Norm}\right)$.}
\end{figure}

The normalized P-field sum at various radii $R$ of $FL_1$ illumination is given by
\begin{equation}
    \label{eq:P_sum3} 
    \mathcal{P}_\mathrm{Norm}(R) = \frac{|\mathcal{P}_\mathrm{Sum}(R)|}{{|\mathcal{P}_\mathrm{Sum}|}_\mathrm{Max}}.
\end{equation}
We plot the theoretically expected values of $\mathcal{P}_\mathrm{Norm}(R)$ in Fig.~\ref{fig:Fres_AC} for different null-to-null beam illumination radius values at $FL_1$ for an initial Gaussian illumination of $1/e^2$ radius $w_0 = \SI{84}{cm}$ with the iris fully open. Also in Fig.~\ref{fig:Fres_AC}, we plot the normalized values of the measured magnitudes of the P-field sum for different $FL_1$ illumination radii. For our measurements, the size of the iris opening was altered and the corresponding illumination radius $R$ on $FL_1$ was measured with a long graduated scale. Then, for each such iris setting, we recorded several $|\mathcal{P}_\mathrm{Sum}|$ measurements with the spectrum analyzer and averaged these measurements to reduce possible noise arising from spurious signal fluctuations. Comparing the theoretical curve to the experimental data, we observe an excellent agreement between the two -- especially given the fact that Fresnel lenses impart significant aberrations to propagating wavefronts. The maximum and minimum experimentally measured values of $\mathcal{P}_\mathrm{Norm}$ are observed at almost the correct values of $R$ which were predicted theoretically. Also, the overall trend of the measured data follows closely the theoretical predictions.

In order to demonstrate that the expected and measured trend of P-field measurement does not simply follow the E-field trend for different values of the beam illumination radius $R$, we also measure the incident average optical power $P_\mathrm{DC}$ at $I$. This was achieved by placing a beam splitter BS before the AC-coupled Menlo-systems fast photo-detector and measuring the average optical power with the aforementioned power meter. These optical power measurements normalized to the maximum value $P_\mathrm{DC-Max}$ of $P_\mathrm{DC}$ (obtained when $FL_1$ was fully illuminated) are given by
\begin{equation}
    \label{eq:norm_dc}
    P_\mathrm{DC-Norm} = \frac{P_\mathrm{DC}}{P_\mathrm{DC-Max}}.
\end{equation}
and are plotted in Fig.~\ref{fig:Fres_DC} which also includes a plot of the theoretically expected normalized average optical power values $\left(P_\mathrm{DC-Norm}\right)_\mathrm{Theoretical}$ for different values of $R$ given by
\begin{equation}
    \label{eq:norm_dc_theor}
    \left(P_\mathrm{DC-Norm}\right)_\mathrm{Theoretical} = 1 - \exp\left( \frac{-2R^2}{w_0^2} \right).
\end{equation}
In Fig.~\ref{fig:Fres_AC}, the experimental measurements follow the theoretical curve very closely for all illumination radius values of $R$. We also show that while $R$ is increased, the average optical irradiance (proportional to E-field sum) increases as more photons are allowed to propagate through the system when the iris clear aperture size is increased. This is not true for the sum of P-fields which initially rises with an increasing $R$, but then reduces and reaches a minimum value before rising again. It also has to be noted that almost perfect P-field cancellation could not be achieved in our experiment as this is only possible for a the case of uniform illumination of Fresnel lenses which is challenging to achieve for such large Fresnel lens aperture sizes in a laboratory environment.

% --------------------------
\section{Conclusion}
\label{sec:conclusions}
% --------------------------

In this paper, we provide experimental validation of the properties of the envelopes of an amplitude modulated optical carrier which we refer to as Phasor fields (or P-fields). We show that these P-fields exhibit wavelike properties of their own and P-field contributions from apertures can be summed in a Huygens-like formulation which is used for the sum of electric fields (or E-fields). We show that these wave-like properties of P-fields allow us to perform non-line-of-sight imaging in the realm of P-fields just like conventional line-of-sight imaging is explained in the realm of E-fields through the Huygens integral. This analogy between imaging with E-fields and P-fields also allows us to use our existing knowledge in LOS imaging to explain and model NLOS imaging. In our experiments, we show that it is possible to achieve 1) a P-field fringe pattern from a rough double slit aperture, 2) P-field focus with no E-field focus and 3) E-field (optical) focus with minimal P-field focus. There is excellent agreement between theory and experimental measurements in all three experiments. The fact that we obtain a P-field airy-disk-like pattern with a P-field lens also allows us to use this physical quantity of P-field sum in future NLOS imaging systems with a minimal post-processing overhead.

% --------------------------
\section*{Funding}
% --------------------------
This work was supported by DARPA REVEAL Program (Grant no. DARPA-BAA-15-44 MSN189781), NASA NIAC Program (Grant no. NNH15ZOA001N-15NIAC A2) AFRSO (Grant no. AFOSR-2014-0003-cidYIP-2015), ONR (Grant no. Open BAA 15-001).
% OSA participates in \href{https://www.crossref.org/fundingdata/}{Crossref's Funding Data}, a service that provides a standard way to report funding sources for published scholarly research. To ensure consistency, please enter any funding agencies and contract numbers from the Funding section in Prism during submission or revisions.

% --------------------------
\section*{Acknowledgments}
% --------------------------
The authors would like to acknowledge Mohit Gupta and his ``Wision Lab'' at the University of Wisconsin -- Madison for lending us their laboratory equipment useful to carry on our experiments.

% --------------------------
\section*{Disclosures}
% --------------------------
The authors declare that there are no conflicts of interest related to this article.

%%%%%%%%%%%%%%%%%%%%%%% References %%%%%%%%%%%%%%%%%%%%%%%%%
\bibliography{biblio}

\begin{thebibliography}{10}
\newcommand{\enquote}[1]{``#1''}

\bibitem{Velten_12}
A.~Velten, T.~Willwacher, O.~Gupta, A.~Veeraraghavan, M.~G. Bawendi, and
  R.~Raskar, \enquote{Recovering three-dimensional shape around a corner using
  ultrafast time-of-flight imaging,} {\protect\JournalTitle{{{Nature
  Communications}}}} \textbf{3}, 745 (2012).

\bibitem{gupta_12}
O.~Gupta, T.~Willwacher, A.~Velten, A.~Veeraraghavan, and R.~Raskar,
  \enquote{Reconstruction of hidden {{3D}} shapes using diffuse reflections.}
  {\protect\JournalTitle{{{Optics Express}}}} \textbf{20}, 19096--108 (2012).

\bibitem{OToole_18}
M.~O'Toole, D.~B. Lindell, and G.~Wetzstein, \enquote{Confocal
  non-line-of-sight imaging based on the light-cone transform,}
  {\protect\JournalTitle{Nature}} \textbf{555}, 338 (2018).

\bibitem{heide_17}
F.~Heide, M.~O'Toole, K.~Zang, D.~Lindell, S.~Diamond, and G.~Wetzstein,
  \enquote{Non-line-of-sight imaging with partial occluders and surface
  normals,} {\protect\JournalTitle{arXiv preprint arXiv:1711.07134}}  (2017).

\bibitem{iseringhausen_18}
J.~Iseringhausen and M.~B. Hullin, \enquote{Non-line-of-sight reconstruction
  using efficient transient rendering,} {\protect\JournalTitle{arXiv preprint
  arXiv:1809.08044}}  (2018).

\bibitem{liu_18_virtual}
X.~Liu, I.~Guill{\'e}n, M.~La~Manna, J.~H. Nam, S.~A. Reza, T.~H. Le,
  D.~Gutierrez, A.~Jarabo, and A.~Velten, \enquote{Virtual wave optics for
  non-line-of-sight imaging,} {\protect\JournalTitle{arXiv preprint
  arXiv:1810.07535}}  (2018).

\bibitem{gupta_15}
M.~Gupta, S.~K. Nayar, M.~B. Hullin, and J.~Martin, \enquote{Phasor imaging: A
  generalization of correlation-based time-of-flight imaging,}
  {\protect\JournalTitle{ACM Transactions on Graphics (ToG)}} \textbf{34}, 156
  (2015).

\bibitem{lin_15}
B.~Lin, A.~R. Nehrir, F.~W. Harrison, E.~V. Browell, S.~Ismail, M.~D. Obland,
  J.~Campbell, J.~Dobler, B.~Meadows, T.-F. Fan \emph{et~al.},
  \enquote{Atmospheric {{CO}}\textsubscript{2} column measurements in cloudy
  conditions using intensity-modulated continuous-wave {{LiDAR}} at 1.57
  micron,} {\protect\JournalTitle{Optics express}} \textbf{23}, A582--A593
  (2015).

\bibitem{gao_12}
S.~Gao and R.~Hui, \enquote{Frequency-modulated continuous-wave lidar using i/q
  modulator for simplified heterodyne detection,} {\protect\JournalTitle{Optics
  letters}} \textbf{37}, 2022--2024 (2012).

\bibitem{lum_18}
D.~J. Lum, S.~H. Knarr, and J.~C. Howell, \enquote{Frequency-modulated
  continuous-wave {{LiDAR}} compressive depth-mapping,}
  {\protect\JournalTitle{Optics express}} \textbf{26}, 15420--15435 (2018).

\bibitem{Heide_14}
F.~Heide, L.~Xiao, W.~Heidrich, and M.~B. Hullin, \enquote{Diffuse mirrors:
  {{3D}} reconstruction from diffuse indirect illumination using inexpensive
  time-of-flight sensors,} in \emph{Proceedings of the IEEE Conference on
  Computer Vision and Pattern Recognition,}  (2014), pp. 3222--3229.

\bibitem{kadambi_16}
A.~Kadambi, H.~Zhao, B.~Shi, and R.~Raskar, \enquote{Occluded imaging with
  time-of-flight sensors,} {\protect\JournalTitle{ACM Transactions on Graphics
  (ToG)}} \textbf{35}, 15 (2016).

\bibitem{Reza18j}
S.~A. Reza, M.~{La Manna}, and A.~Velten, \enquote{A physical light transport
  model for non-line-of-sight sight imaging applications,}
  {\protect\JournalTitle{arXiv preprint arXiv:1802.01823}}  (2018).

\bibitem{Reza18c}
S.~A. Reza, M.~{La Manna}, and A.~Velten, \enquote{Imaging with phasor fields
  for non-line-of sight applications,} in \emph{Imaging and Applied Optics 2018
  (3D, AO, AIO, COSI, DH, IS, LACSEA, LS\&C, MATH, pcAOP),}  (Optical Society
  of America, 2018), p. CM2E.7.

\bibitem{goodman2005introduction}
J.~W. Goodman, \emph{Introduction to Fourier optics} (Roberts and Company
  Publishers, 2005).

\end{thebibliography}

\end{document}